\documentclass[a4paper,10pt]{article}
 \usepackage{fullpage}
 \usepackage{setspace}
 \usepackage{subeqn}
 \usepackage{graphicx}
\begin{document}
 
 \begin{center}
 \large{\textbf{Liquid Water Oceans in Ice Giants}}\\
  \end{center}
 
 \begin{center}
Sloane J. Wiktorowicz* and Andrew P. Ingersoll\\
Division of Geological and Planetary Sciences, California Institute of Technology, Pasadena, CA 91125\\
*Corresponding author email address: sloane@gps.caltech.edu\\
  \end{center}
  
  \smallskip
  \begin{center}
\textbf{ABSTRACT}\\
 \end{center}
   \bigskip
 
Aptly named, ice giants such as Uranus and Neptune contain significant amounts of water. While this water cannot be present near the cloud tops, it must be abundant in the deep interior. We investigate the likelihood of a liquid water ocean existing in the hydrogen-rich region between the cloud tops and deep interior. Starting from an assumed temperature at a given upper tropospheric pressure (the photosphere), we follow a moist adiabat downward. The mixing ratio of water to hydrogen in the gas phase is small in the photosphere and increases with depth. The mixing ratio in the condensed phase is near unity in the photosphere and decreases with depth; this gives two possible outcomes. If at some pressure level the mixing ratio of water in the gas phase is equal to that in the deep interior, then that level is the cloud base. The gas below the cloud base has constant mixing ratio. Alternately, if the mixing ratio of water in the condensed phase reaches that in the deep interior, then the surface of a liquid ocean will occur. Below this ocean surface, the mixing ratio of water will be constant. A cloud base occurs when the photospheric temperature is high. For a family of ice giants with different photospheric temperatures, the cooler ice giants will have warmer cloud bases. For an ice giant with a cool enough photospheric temperature, the cloud base will exist at the critical temperature. For still cooler ice giants, ocean surfaces will result. A high mixing ratio of water in the deep interior favors a liquid ocean. We find that Neptune is both too warm (photospheric temperature too high) and too dry (mixing ratio of water in the deep interior too low) for liquid oceans to exist at present. To have a liquid ocean, Neptune's deep interior water to gas ratio would have to be higher than current models allow, and the density at 19 kbar would have to be $\approx$ 0.8 g/cm$^{3}$. Such a high density is inconsistent with gravitational data obtained during the Voyager flyby. In our model, Neptune's water cloud base occurs around 660 K and 11 kbar, and the density there is consistent with Voyager gravitational data. As Neptune cools, the probability of a liquid ocean increases. Extrasolar Òhot Neptunes,Ó which presumably migrate inward toward their parent stars, cannot harbor liquid water oceans unless they have lost almost all of the hydrogen and helium from their deep interiors.

\bigskip
\begin{center}
\bf{1. INTRODUCTION}
\end{center}
\bigskip

Water is spectroscopically undetectable in both Uranus and Neptune; its saturated mixing ratio in their cold photospheres is less than 10$^{-25}$. However, the compressed, deep interior density of these planets is strikingly close to that of pure water (Hubbard 1999). Additionally, successful density models require an ice to rock mass ratio greater than unity (Hubbard et al. 1995) and gas comprising less than 18\% of the planet's mass (Podolak et al. 2000). When addressing planetary mass, we take ÒgasÓ to mean hydrogen and helium, and we assume ÒiceÓ consists of water ice as well as methane, ammonia, and hydrogen sulfide. There is evidence based on CO observations that the enhancement of oxygen in Neptune's atmosphere with respect to the solar value is larger than that for carbon, nitrogen, and sulfur (Lodders and Fegley 1994). Since oxygen is the most abundant element in the solar system next to hydrogen and helium, water is thought to be the dominant component of ice in the outer solar system. Thus, there must be a significant water reservoir in Uranus and Neptune. Indeed, there has been speculation about ÒoceansÓ in their deep interiors (Atreya 1986, p. 64; Hubbard et al. 1995), but these ÒoceansÓ describe ionic phase transitions at thousands of degrees Kelvin.

We explore the necessary conditions for bona fide liquid water-hydrogen oceans to exist in the upper interior of Neptune, where pressure is less than about 20 kbar and temperature is less than about 800 K. We define the word ÒoceanÓ to mean a body with an interface between a hydrogen-rich, saturated vapor and a water-rich, liquid ocean. We set up favorable conditions for an extant Neptunian water ocean to show that its existence is unlikely: water, hydrogen, and helium are assumed to be well mixed in the interior of the planet, and we assume a saturated (moist) water adiabat descends from the photosphere. The photospheric adiabat and the interior adiabat join at a phase boundary that is either a cloud base or an ocean surface. The temperature-entropy diagram for pure water (Fig. 1) provides a qualitative illustration of how this works. We show temperature decreasing upward in this figure, so an adiabat from the deep interior is a vertical line from below. If it approaches the phase boundary on the right (the high-entropy side of the critical point), then liquid droplets will form in the vapor. If the interior adiabat approaches the phase boundary on the left (the low-entropy side of the critical point), then vapor bubbles will form in the liquid. The former is analogous to a cloud base and the latter is analogous to an ocean surface. The critical adiabat is the one that intersects the phase boundary at the critical point, which is also the warmest point (647 K) on the phase boundary.

Since hydrogen is present in ice giants in addition to water vapor, the critical point at each pressure level will depend on its composition. To describe these mixtures, Fig. 1 should be three-dimensional with composition as the third axis. The critical point from Fig. 1 then becomes a critical curve. The locations of phase boundaries will be strongly affected by mixture composition. Seward and Franck (1981), hereafter referred to as SF, experimentally identify the critical curve. They also investigate the phase boundaries of water-hydrogen mixtures for temperatures, pressures, and compositions below 654 K, 2.5 kbar, and 60 mol-\% hydrogen to (water + hydrogen).

In an ice giant with a moist adiabat connecting the photosphere to a phase transition, the existence of a cloud base or an ocean surface depends on two input parameters: the water to total gas mixing ratio of the deep interior and the photospheric temperature. The moist adiabat extending down from the photosphere contains two phases, a gas phase and a condensed phase. A cloud base will result if the gas phase reaches the deep interior mixing ratio before the condensed phase does. Conversely, an ocean surface will result if the condensed phase reaches the deep interior mixing ratio before the gas phase does.

To understand liquid water oceans on Neptune, it is helpful to consider an idealized model of the Earth's ocean/atmosphere system. In equilibrium, the concentration of dry air (mostly nitrogen, oxygen, and argon) in the ocean is set by its solubility and is denoted by X$_{dry}$. Under present conditions, X$_{dry}$ is $\approx$ 2 x 10$^{-5}$ by mass. We define the photospheric temperature T$_{phot}$ as the atmospheric temperature at the 0.4 bar level. The atmospheric temperature and pressure follow a pseudo-adiabat Ð a moist adiabatic expansion in which the condensed water is removed from the system as soon as it forms (e.g., Emanuel 1994; Salby 1996). Although the condensate is removed, the atmosphere is saturated at every pressure level, which means that a liquid water droplet suspended in the atmosphere just above the surface has the same X$_{dry}$ as the ocean itself. With these assumptions, the values of X$_{dry}$ and T$_{phot}$ determine everything about the system, including the ocean temperature T$_{ocean}$ and the partial pressures of water and dry air at the ocean interface. If T$_{phot}$ were to increase, the mixing ratio of water on the moist adiabat would increase. Then, for X$_{dry}$ fixed, T$_{ocean}$ would have to increase in order to match the increased mixing ratio of water in the atmosphere (the Henry's Law constant for air is a weaker function of temperature than the vapor pressure of water is). Increasing T$_{ocean}$ is like moving toward the critical point from the left (low temperature) side of Fig. 1. Alternately, for T$_{ocean}$ fixed, X$_{dry}$ would have to decrease. This lowers the equilibrium partial pressure of dry air relative to water and again matches the increased mixing ratio of water in the atmosphere.

In temperature-composition space, the boundary between the region of cloud base solutions and the region of ocean surface solutions will be a line. We refer to this boundary as the critical ocean, and it lies at the critical temperature for its composition. At the critical ocean, both the water mixing ratios and densities will be equal between the gas and condensed phases. As in Fig. 1, no phase transitions are possible for temperatures higher than the critical temperature.

For a suite of ice giants with different photospheric temperatures, different deep interior water mixing ratios, and different atmospheric gas masses, the qualitative effects on the existence of oceans are as follows. A cooler photosphere results in a photospheric adiabat with decreased water mixing ratio in the gas phase (and therefore increased water mixing ratio in the condensed phase) at each pressure level. Since the condensed phase mixing ratio will reach the deep interior value before the gas phase ratio does, the photospheric adiabat will terminate in an ocean surface. For an ice giant with a large water mixing ratio in the interior, the condensed phase ratio will again reach the interior value before the gas phase ratio does. This ice giant will also contain an ocean surface. As in Fig. 1, higher-entropy photospheric adiabats terminate in cloud bases. Entropy can be increased either by increasing the photospheric temperature or by decreasing the pressure at a given temperature. The latter is similar to decreasing the atmospheric gas mass.

Even though we complicate the Neptune calculations by assuming van der Waals gases as well as condensation of methane, ammonia, and hydrogen sulfide, the qualitative aspects derived above still apply. We pin the moist adiabat at 59 K and 0.4 bar (see Fig. 8 in Burgdorf et al. 2003), and we extend it downward until a phase transition is reached. We determine which of these transitions is likely by following the photospheric adiabat until it intersects the phase transition curves of SF. Published models of Neptune's density structure are then compared to our density estimates.

\bigskip
\begin{center}
\bf{2. PHOTOSPHERIC ADIABAT}\\
\bf{2.1. Construction}
\end{center}
\bigskip

Because gas at the temperature and pressure of a phase transition is non-ideal, the van der Waals relation is the basis of our model:

\begin{equation}
P=\frac{RT}{V-b_{H2}}-\frac{a_{H2}}{V^2},\\
\end{equation}

\bigskip
\noindent where a$_{H2}$ and b$_{H2}$ are the molar van der Waals coefficients for hydrogen (2.45$\times$10$^{11}$ erg cm$^{3}$ mol$^{-2}$ and 26.61 cm$^{3}$ mol$^{-1}$, respectively; see Fishbane et al. (2005)), T is the temperature, and V is the molar volume of the water-hydrogen mixture. Quantities without subscripts, other than P, R, T, or V, denote water vapor.

We assume that internal heat is convected up from the deep interior of ice giants to the photosphere, at about 0.4 bar, and then radiated to space. Thus, it is reasonable to assume their pressure-temperature profiles follow adiabats. Hydrogen and helium affect molecular mass and heat capacity along the photospheric adiabat. Carbon, nitrogen, and sulfur are assumed to exist as methane, ammonia, and hydrogen sulfide. We include their effect on molecular mass and heat capacity in addition to allowing their condensation. Thus, we assume that these species are also saturated along the photospheric adiabat. Since we do not take into account the heat capacity or volume of the condensed phases, the photospheric, moist adiabat is similar to a pseudoadiabat.

The temperature versus molar volume profile is calculated from the following equation, which is derived in the Appendix:

\begin{equation}
\frac{dV}{dT}=-\left[\displaystyle\frac{C_v+\displaystyle\sum_{i}\left(\frac{L_i}{1+q_i}\right)\left(\frac{\partial q_i}{\partial T}\right)_V}{\displaystyle\frac{RT}{V-b_{H2}}+\displaystyle\sum_{i}\left(\frac{L_i}{1+q_i}\right)\left(\frac{\partial q_i}{\partial V}\right)_T}\right].\\
\end{equation}

\bigskip
\noindent The summations occur over each species i (hydrogen, helium, water vapor, methane, ammonia, and hydrogen sulfide). Since hydrogen and helium are not condensing, their latent heats are set to zero. R is the molar gas constant and q$_{i}$ is the molar mixing ratio of the condensing species to the other five species, given by the following:

\begin{equation}
q_i=\displaystyle\frac{f_i}{1-f_i},\\
\end{equation}
 
 \bigskip
\noindent where f$_i$ is the molar mixing ratio of species i to total gas and L$_i$ is the species' latent heat (by mole) of sublimation or condensation. The mixture's molar heat capacity at constant volume is given by C$_v$, which is derived in the Appendix:

\begin{equation}
C_v=\displaystyle\sum_{i}\left(C_{p_i}-\displaystyle\frac{R}{1-\displaystyle\frac{2a_{H2}\left(V-b_{H2}\right)^2}{RTV^3}}\right)f_i\\
\end{equation}
 
 \bigskip
\noindent for a van der Waals gas. Here, C$_p$ is the mixture's molar heat capacity at constant pressure (the weighted mean, by mole, of the heat capacities of all six species).

For each species, the C$_p$ values and latent heats of sublimation are taken from Atreya (1986). The latent heats of vaporization for methane and ammonia are 8.519$\times$10$^{10}$ erg mol$^{-1}$ for T $\ge$ 90.6 K (National Institute of Standards and Technology) and 2.5$\times$10$^{11}$ erg mol$^{-1}$ for T $\ge$ 194.95 K (Osborne and van Dusen 1918), respectively. The temperature dependences of L and the saturation vapor pressure e for pure water, are taken from Wagner and Pruss (1993). Saturation vapor pressures versus temperature for methane (Ziegler 1959), hydrogen sulfide (Giauque and Blue 1936; Vorholz et al. 2002), and ammonia (Karwat 1924; International Critical Tables 1928) are compiled in Atreya (1986). Latent heat and saturation vapor pressure of pure water are used because we do not have an adequate description of how these quantities vary, as a function of water vapor mixing ratio, in a water-hydrogen mixture. Limited data on saturation vapor pressure are indeed given in SF, so an interpolation is required to integrate the photospheric adiabat. The accuracy of this interpolation will be addressed later (Fig. 3). As will be seen in Section 4.1, the fact that latent heat goes to zero at 647 K (the critical point for pure water) causes unphysical behavior at higher temperatures. We attempt to remedy this by extrapolating data from T $<$ 600 K to predict high temperature behavior.

Temperature and volume along the photospheric adiabat translate to pressure according to the van der Waals relation (Eq. 1). In order to determine f, the mixing ratio of water to total gas, we modify Eq. 1 to describe water vapor in the gas phase: P becomes e, a$_{H2}$ and b$_{H2}$ become the coefficients for water (a and b), and V becomes V/f. This is because the volume per mole of water is the volume per mole of the mixture (V) times the moles of mixture per mole of water (1/f). Therefore,
 
 \begin{equation}
 e=\displaystyle\frac{RTf}{V-bf}-\displaystyle\frac{af^2}{V^2}.\\
 \end{equation}
 
 \bigskip
 \noindent Here, a = 5.507$\times$10$^{12}$ erg cm$^{3}$ mol$^{-2}$ and b = 30.4 cm$^{3}$ mol$^{-1}$ (Fishbane et al. 2005). Using Eq. 5, we find
 
 \begin{subequations}
 \begin{equation}
 \left(\displaystyle\frac{\partial f}{\partial T}\right)_V=\displaystyle\frac{V^2\left(V-bf\right)^2\displaystyle\frac{de}{dT}-RV^2f\left(V-bf\right)}{RTV^3-2af\left(V-bf\right)^2}\\
 \end{equation}
 
 \begin{equation}
 \left(\displaystyle\frac{\partial f}{\partial V}\right)_T=\displaystyle\frac{RTV^3f-2af^2\left(V-bf\right)^2}{RTV^4-2Vaf\left(V-bf\right)^2}.\\
 \end{equation}
 \end{subequations}

\bigskip
\noindent From the definition of q in Eq. 3, we can determine the quantities \begin{math} \left(\displaystyle\frac{\partial q}{\partial T}\right)_V \end{math} and \begin{math} \left(\displaystyle\frac{\partial q}{\partial V}\right)_T \end{math}, which are necessary for Eq. 2.

In ice giants, the other condensable species form cloud bases above the level of significant water condensation, and their main effect is to lower the temperature of the photospheric adiabat within the water cloud. To make the calculations simpler, we use the ideal gas approximation for the other condensable gases. The van der Waals constants in Eq. 5 can be set to zero, and
 
 \begin{equation}
 e_j=\displaystyle\frac{RTf_j}{V}.\\
 \end{equation}
 
 \bigskip
 \noindent Here, the summation index j is taken to mean methane, ammonia, and hydrogen sulfide. For numerical integration, it is useful to formulate \begin{math} \left(\displaystyle\frac{\partial f_j}{\partial T}\right)_V \end{math} and \begin{math} \left(\displaystyle\frac{\partial f_j}{\partial V}\right)_T \end{math} in terms of f$_j$:
 
 \begin{subequations}
 \begin{equation}
 \left(\displaystyle\frac{\partial f_j}{\partial T}\right)_V=\left(\displaystyle\frac{1}{e_j}\displaystyle\frac{de_j}{dT}-\displaystyle\frac{1}{T}\right)f_j,\\
 \end{equation}
 
 \begin{equation}
 \left(\displaystyle\frac{\partial f_j}{\partial V}\right)_T=\displaystyle\frac{f_j}{V}.\\
 \end{equation}
 \end{subequations}
 
 \bigskip
 \begin{center}
 \bf{2.2 Application to Neptune}
 \end{center}
 \bigskip
 
Neptune itself might not be fully mixed from photosphere to rocky core. It is possible that there exists stable stratification in the deep interior. However, the maximum water mixing ratio along the photospheric adiabat cannot exceed the deep interior value, because density must increase with depth. Thus, the true water mixing ratio at the phase transition must be less than or equal to the deep interior value. To provide our Neptune model with the most optimistic parameters for the existence of oceans, we assume the phase transition occurs when the water mixing ratio equals the deep interior ratio.

We tie the van der Waals, photospheric adiabat to a pressure-temperature estimate (59 K at 0.4 bar) obtained by a combination of Voyager radio occultation experiments (Lindal 1992) and Infrared Space Observatory observations (Burgdorf et al. 2003). We hold Neptune's helium to hydrogen gas mole fraction at 19/81 along the photospheric adiabat and also in the deep interior (Lindal 1992). Baines et al. (1995) find a constant methane mixing ratio below the methane cloud base of \begin{math} f_{CH4}=2.2^{+0.5}_{-0.6} \end{math} mol-$\%$. We estimate the mixing ratio of the other condensables at their cloud bases by assuming solar values of [C]:[N]:[S] along the photospheric adiabat. Given \begin{math} \frac{[C]}{[H_2]}=7.96\times10^{-4}, \frac{[N]}{[H_2]}=2.24\times10^{-4}, \end{math} and \begin{math} \frac{[S]}{[H_2]}=3.7\times10^{-5} \end{math} in the Sun (Gautier et al. 1995), the ammonia and hydrogen sulfide cloud bases occur when \begin{math} f_{NH3}=0.6^{+0.1}_{-0.2} \end{math} mol-$\%$ and \begin{math} f_{H2S}=0.10^{+0.02}_{-0.03} \end{math} mol-$\%$.

We overlay the photospheric adiabat on the phase transition curves from Figure 2 of SF, and we present them as Fig. 2. The thick line is the photospheric adiabat, and the thin lines are the curves from SF. Each pressure-temperature point between 450 K and 650 K has two phase transition curves passing through it. Each curve is labeled with its corresponding percentage hydrogen to (water + hydrogen) molar mixing ratio, X$_{H2}$. The curves with high values of X$_{H2}$ give the composition of the vapor, and the curves with low values of X$_{H2}$ give the composition of the liquid when the two phases are in equilibrium. A critical point occurs when the liquid and vapor have the same composition, i.e., where the curve for a given composition has infinite dP/dT. It is important to note that liquid water infused with more than about 1 mol-$\%$ hydrogen is in equilibrium with vapor only at a higher pressure than the critical pressure for that mixture. For a given composition, then, an ocean surface must lie at supercritical pressure.

To estimate the uncertainty in the photospheric adiabat, dV/dT from Eq. 2, we compare experimentally determined (P, T, f) data to those predicted along the photospheric adiabat. We follow two saturated vapor to dry vapor phase boundaries of SF which correspond to f = 10 mol-$\%$ and 40 mol-$\%$ (X$_{H2}$ = 90 mol-$\%$ and 60 mol-$\%$, respectively). Note that f = X = 100 mol-$\%$ - X$_{H2}$ for SF data because their system only contains water and hydrogen. Also note that their phase boundaries are only printed for T $>$ 450 K. Since we know T and f from SF (f is either 10 mol-$\%$ or 40 mol-$\%$), we predict the volume along the photospheric adiabat by solving for V in Eq. 5 (saturation vapor pressure, e, is only a function of T). Using P and T from SF, we calculate the volume along their phase boundaries, VSF, by solving for V in Eq. 1. The quantity V$_{SF}$/V is a measure of the discrepancy between the photospheric adiabat and experimental data.

Fig. 3 shows V$_{SF}$/V for nine pressure-temperature points in the temperature range 450 K $<$ T $<$ 630 K. The conditions for T $\le$ 273 along the photospheric adiabat are ideal: \begin{math} \left|\displaystyle\frac{PV}{RT}-1\right|=7.1\% \end{math} at 273 K. Since the van der Waals equation of state tends towards the ideal gas formulation at low temperatures and pressures, we assume that V$_{SF}$/V = 1 for T $\le$ 273 along the photospheric adiabat. By multiplying V$_{SF}$/V by the calculated dV/dT from Eq. 2, we can correct the photospheric adiabat to agree with SF.

\begin{equation}
\displaystyle\frac{dV}{dT}_{true}=\displaystyle\frac{V_{SF}}{V}\displaystyle\frac{dV}{dT}
\end{equation}

\bigskip
\noindent Therefore, we fit a fourth-order polynomial to V$_{SF}$/V versus temperature from the nine pressure-temperature points to provide a correction factor at each temperature. We force the value and slope of this factor at T = 273 K to be one and zero, respectively. The dotted lines in Fig. 3 represent the 1$\sigma$ error bounds on the fourth-order fit. The corrected, photospheric adiabat, extending from 59 K to the critical temperature of pure water (647 K), is shown as the middle, thick curve in Fig. 4. The 1$\sigma$ upper and lower bounds to the fourth-order fit are multiplied by dV/dT from Eq. 2 to determine the 1$\sigma$ upper and lower bounds to the photospheric adiabat. These error bounds are given as the thin curves in Fig. 4.

\bigskip
\begin{center}
\bf{3. PHASE TRANSITION}\\
\bf{3.1. Cloud base}
\end{center}
\bigskip

By assuming values for both Neptune's photospheric temperature and its deep interior water mixing ratio, we determine which phase transition exists. Therefore, as we integrate downward from the 59 K photosphere, the target is the deep interior mixing ratio. We estimate this value by assuming Neptune has a deep interior ice to rock mass ratio of 3.0$^{+0.5}_{-2.0}$ : 1 (Podolak and Reynolds 1984; Podolak et al. 1991) and a gas mass of 2.0$^{+1.2}_{-0.5}$ M$_\oplus$ (Gudkova et al. 1988; Hubbard et al. 1995; Podolak et al. 2000). The above authors obtained these values by fitting density models to Neptune's gravitational harmonics which were measured during the Voyager flyby. We assume that the deep interior [H$_2$]:[He] value is the same as the value in the atmosphere (19/81). This will comprise the deep interior gas mass given above. The mass and mole fraction makeup of Neptune's deep interior gas is shown in Table 1.

Since the total planetary mass is 17.14 M$_\oplus$ (Hubbard et al. 1995), the corresponding deep interior ice mass is 11.4$^{+0.6}_{-2.1}$ M$_\oplus$. Since the makeup of this ice is unknown, we assume that the ratios [C]:[N]:[O]:[S] are solar in the deep interior. The corresponding mass and mole fraction components of Neptune's deep interior ice are given in Table 1.

As can be seen from the bold value in Table 1, the deep interior water vapor to total gas mixing ratio (hydrogen, helium, water, methane, ammonia, hydrogen sulfide) is \begin{math} f_{interior}=26.9^{+5.2}_{-9.5} \end{math} mol-$\%$. This means that a water vapor cloud base will be reached if the water vapor mixing ratio f reaches f$_{interior}$ along the photospheric adiabat before the critical temperature is reached (see Section 1). We have a nominal, deep interior water mixing ratio and its associated upper and lower bounds, and we also have a nominal, photospheric adiabat with upper and lower bounds (see Fig. 4). The combination that favors a cloud base is the low-pressure, photospheric adiabat bound paired with the lower mixing ratio bound (f$_{interior}$ = 17.4 mol-$\%$), and it reaches cloud base at 623 K and 5.0 kbar. The combination that favors a liquid ocean is the high-pressure, photospheric adiabat bound paired with the upper mixing ratio bound (f$_{interior}$ = 32.1 mol-$\%$), and a cloud base is reached at 705 K and 19.5 kbar. The nominal, photospheric adiabat paired with the nominal interior mixing ratio (f$_{interior}$ = 26.9 mol-$\%$) is our best estimate, and it reaches cloud base at 663 K and 10.7 kbar. Therefore, the water vapor cloud base is reached at T = 663$^{+42}_{-41}$ K and P = 10.7$^{+8.8}_{-5.7}$ kbar. These temperatures are above 647 K, which is the critical temperature for pure water, because the mixture consists of water and hydrogen. This will be discussed in Section 4.1.

 \bigskip
 \begin{center}
 \bf{3.2 Supercritical fluid}
 \end{center}
 \bigskip
 
Below the cloud base, the atmosphere conforms to a dry (non-condensing), adiabatic gas with about 27 mol-$\%$ water vapor to total gas. The pressure-temperature profile of this dry adiabat can be found by setting L = 0 in Eq. 2, keeping df/dT = 0, and solving for pressure in Eq. 1. Eventually, as one descends further, the gas will slowly transition into a supercritical fluid whose density equals that of a liquid of the same composition. This supercritical fluid is not a true ocean with a saturated vapor to liquid interface. Hot, ionic ÒoceansÓ have been predicted in Neptune's deep interior (Atreya 1986, p. 64; Hubbard et al. 1995), and they would lie at $\approx$ 2,000 K (Atreya, et al. 2005).

\pagebreak
\bigskip
\begin{center}
\bf{4. OCEAN REQUIREMENTS}\\
\bf{4.1. Deep interior mixing ratio}
\end{center}
\bigskip

By assuming a value for Neptune's photospheric temperature and by leaving the deep interior water mixing ratio as a free parameter, we find the minimum deep interior water mixing ratio that will allow an ocean to exist. This particular ocean will be a critical ocean. A critical ocean will occur if the photospheric adiabat (in temperature-composition space) intersects the critical curve, and if the pressure at that intersection is higher than the critical pressure for that composition (see Fig. 2 and the text in Section 2.2 describing it). The composition at this intersection is the minimum deep interior mixing ratio that allows an ocean to exist. Since we only aim to examine the requirements for a critical ocean, it is unnecessary to calculate the water mixing ratio in the condensed phase. To determine the location of a cool ocean, however, the water mixing ratio in the condensed phase must be calculated. This will require the mixing ratios of the other condensables to be determined in the water-hydrogen condensate. Currently, experimental data are not sufficient for this to be done accurately.

We extrapolate the data in Table 1 of SF (and the additional data point on page 3, column 2 of their paper) to construct the critical curve over a large temperature range, and thus to allow the photospheric adiabat to intersect at high temperature oceans.  This table lists critical temperature and pressure for a variety of compositions. Because the photospheric adiabat utilizes f (mixing ratio of water to total gas), we employ composition as the water to (water + hydrogen) mixing ratio, defined as X, Thus, X + X$_{H2}$ = 100 mol-$\%$. We fit the data with piecewise cubic Hermite polynomials because a spline fit appears unphysical. The critical curve in temperature-X space is given as Fig. 5, and the critical curve in pressure-X space is given as Fig. 6.

We extrapolate the photospheric adiabat, with a spline in temperature-X space, to temperatures higher than 647 K. This is because water-hydrogen phase transitions can occur at higher temperatures. Unfortunately, using all T $<$ 647 K in this fit gives a multivalued profile. This unphysical behavior occurs because we are forced to approximate the mixture's latent heat and saturation vapor pressure with the expressions given for pure water: this is invalid near 647 K. However, including only T $<$ 600 K in the extrapolation eliminates this problem. It is a reasonable approximation because latent heat is fairly constant with temperature until it very quickly goes to zero near the critical temperature. In Fig. 7, the photospheric adiabat is given as the thick, solid curve, and its extension to T $>$ 647 K is given as the solid curve. The critical curve is shown as the thin, solid curve. Dashed curves indicate the associated 1$\sigma$ errors on each curve. Finally, the vertical, dotted lines show the range of deep interior mixing ratios assumed for Neptune.

We attempt to place upper and lower limits on the shape of the extended, photospheric adiabat, and we acknowledge that it is a crude extrapolation. The extended, photospheric adiabat intersects the critical curve, and therefore terminates in a critical ocean, at T = 702.1$^{+6.1}_{-7.3}$ K and X = 38.8 $\pm$ 1.4 mol-$\%$. When factoring in the mixing ratios of the other species, we find f = 33.0 $\pm$ 1.3 mol-$\%$. By extending the photospheric adiabat in pressure-temperature space, we determine the pressure of this critical ocean to be P = 18.7$^{+6.6}_{-5.3}$ kbar. Since the critical pressure for this composition is lower than the pressure of the critical ocean (P = 11.65$^{+0.21}_{-0.42}$ kbar, see Fig. 6), we verify that the critical ocean is indeed liquid. An extant ocean in Neptune thus requires a deep interior water mixing ratio of at least f = 33.0 $\pm$ 1.3 mol-$\%$, but we estimate its current value to be f$_{interior}$ = 26.9$^{+5.2}_{-9.5}$ mol-$\%$. Neptune is therefore slightly too dry to harbor oceans.

 \bigskip
 \begin{center}
 \bf{4.2 Gravitational signature}
 \end{center}
 \bigskip
 
We investigate whether the photospheric adiabat is consistent with density models of Neptune's interior. At each pressure level, we calculate the density by dividing molar mass by molar volume, V. We treat molar mass as simply a mean, weighted by mixing ratio, of the molar masses of the constituent species. We plot pressure against density along the photospheric adiabat as Fig. 8, and we overlay these results on Fig. 5 from Hubbard et al. (1995). As can be seen, the photospheric adiabat is consistent with density models from photosphere to cloud base. We find the 5 to 20 kbar cloud base from Section 3.1 has a density of 0.221$^{+0.048}_{-0.076}$ g/cm$^3$, while density models predict 0.09 to 0.30 g/cm$^3$. The location of the cloud base is above the density discontinuity at $\approx$ 100 kbar, as expected.

We now determine whether the critical ocean is consistent with density models. To estimate the density of the critical ocean, we ignore the contribution of all species except water and hydrogen. This is because water and hydrogen dominate the vapor, and the dissolved mole fraction of the other species should be even lower in the liquid. We use the following law of additive volumes (Hubbard 1972):

\begin{equation}
\displaystyle\frac{1}{\rho_{mix}}=\displaystyle\sum_{i=1}^{n}\displaystyle\frac{M_i}{\rho_i},\\
\end{equation}

\bigskip
\noindent where M$_i$ and $\rho_i$ are the mass mixing ratio and density, respectively, of species i. To calculate the densities of hydrogen and water at the surface of the 19 kbar critical ocean, we use equation of state fits complied by Hubbard et al. (1995) from various sources. Plugging in the densities for these two species, we find the critical ocean has a density of 0.772$^{+0.061}_{-0.059}$ g/cm$^3$. Based on planetary density models calculated from Voyager gravitational constraints, the density between 13 and 25 kbar (range of pressures at the critical ocean) lies between 0.14 g/cm$^3$ and 0.34 g/cm$^3$ (see Fig. 8), which is inconsistent with an ocean; the required water mixing ratio is too high. Again, Neptune is too dry to harbor oceans.

 \bigskip
 \begin{center}
 \bf{4.3 Photospheric temperature}
 \end{center}
 \bigskip
 
By assuming a value for Neptune's deep interior water mixing ratio and by leaving the photospheric temperature as a free parameter, we aim to find the maximum photospheric temperature that will allow an ocean to exist. Neptune is not in thermal equilibrium with the Sun's radiation; since its thermal emission is 2.6 times as strong as its solar heating (Hubbard et al. 1995), Neptune is slowly cooling. As an adiabatic atmosphere cools, its entropy decreases, which moves the phase boundary to the left in Fig. 1. As an ice giant with a high-entropy, low-temperature cloud base cools, its cloud base will migrate to high temperature. After further cooling of the ice giant, the cloud base will pass through the critical point to become a liquid ocean at high temperature. Finally, the cloud base will transition to a liquid ocean at low temperature. Thus, it is worthwhile to ask whether Neptune will eventually cool enough to permit the existence of liquid water oceans. Since the critical point for hydrogen gas is 33.2 K, 13.0 bar (National Institute of Standards and Technology), hydrogen gas will begin to condense at cooler temperatures. Since we do not take hydrogen condensation into account, we do not attempt to model ice giants cooler than 30 K at 0.4 bar. We investigate the probability of an ocean's existence by considering a suite of ice giants with photospheric temperatures higher than 30 K.

We assume f in the ocean will equal the deep interior mole fraction, f$_{interior}$ = 26.9$^{+5.2}_{-9.5}$ mol-$\%$ (see Section 3.1). Given the atmospheric mixing ratios of the other species (see Section 2.2), X = 32$^{+6}_{-11}$ mol-$\%$ in the ocean. Since the cloud base mixing ratios of the other condensables are much lower than their assumed mixing ratios in the deep interior, X in the ocean will not equal X$_{interior}$. From Figs. 6 and 7, we find the critical point to be at 726$^{+69}_{-28}$ K and 16$^{+12}_{-5}$ kbar for X = 32$^{+6}_{-11}$ mol-$\%$. We set the 0.4 bar temperature to 30 K and integrate the photospheric adiabat, as well as its upper and lower bounds, down to 647 K. We then fit a spline to the temperature-X profile for T $<$ 600 K and extrapolate to higher temperatures. We find the lower, photospheric adiabat bound reaches the lower value of X = 21 mol-$\%$ at 636 K. Since this pressure level is much cooler than its 796 K critical temperature, the lower, photospheric adiabat bound terminates in a cloud base. The nominal, photospheric adiabat reaches the nominal X = 32 mol-$\%$ at 707 K, which is cooler than that pressure level's 726 K critical temperature. Thus, the nominal, photospheric adiabat also intersects a cloud base.

However, the upper, photospheric adiabat bound does not reach the upper value of X = 38 mol-$\%$ before its 698 K critical temperature. In fact, the upper bound to the photospheric adiabat intersects the critical curve at 720.9$^{+6.3}_{-5.9}$ K and X = 33.26$^{+0.54}_{-0.53}$ mol-$\%$, which implies that it reaches a critical ocean. The uncertainties in temperature and composition of this critical ocean are solely due to the uncertainties of SF in measuring the critical curve. Extending the upper, photospheric adiabat bound with a spline fit in pressure-temperature space, we find this critical ocean to lie at 155$^{+14}_{-12}$ kbar. These calculations are probably unqualified to accurately predict a pressure this high. However, this critical ocean certainly lies at a higher-than-critical pressure of 15.1$^{+2.0}_{-1.9}$ kbar. Therefore, we verify that the upper, photospheric adiabat bound (pinned to a 30 K photosphere) intersects a critical ocean.

The existence of a critical ocean under a 30 K photosphere can only occur if the actual profile intersects the critical curve before reaching the deep interior water mixing ratio. Thus the probability of an ocean's existence is related to the probability that Neptune's deep interior water mixing ratio is higher than the nominal value. The minimum value of X necessary for an ocean is the composition at the intersection between the nominal, photospheric adiabat and the critical curve. Assuming Gaussian statistics, this occurs at X = 33.37$^{+0.65}_{-0.64}$ mol-$\%$, which is 0.22 $\pm$ 0.11$\sigma$ away from the ocean's X = 32$^{+6}_{-11}$ mol-$\%$. Therefore, the probability of such a high water mixing ratio, and thus the probability of a 30 K photosphere terminating in an ocean surface, is 41.5 $\pm$ 4.2$\%$. It should be emphasized that the probabilistic nature of an ocean's existence is due to uncertainty both in the thermodynamics of the photospheric adiabat and in the value of Neptune's deep interior water mixing ratio.

This approach can also be applied to a suite of photospheres with higher temperatures, but the probability of an ocean will decrease with increasing photospheric temperature (see Fig. 9). Note that Neptune's current 59 K photosphere has only a 13.1$^{+5.4}_{-4.3}\%$ probability of terminating in an ocean. While a 13$\%$ probability is not insignificant, Voyager gravitational data verify that Neptune has no oceans (see Section 4.2). Thus, Neptune is too warm to harbor oceans.

Will Neptune ever cool down to 30 K? Simply finding the temperature at which Neptune's thermal emission is in equilibrium with solar buffering provides a very rough estimate of the extent to which it can cool:

\begin{equation}
4\pi R^2\sigma T_{e}^{4}=\left(1-\Lambda\right)\pi R^2\left(\pi F_{Sun}\right)\\
\end{equation}

\bigskip
\noindent Here R is planetary radius, $\sigma$ is the Stefan-Boltzmann constant, T$_e$ (effective temperature) is assumed to be the temperature at 0.4 bars, $\Lambda$ is Bond albedo, and $\pi$F$_{Sun}$ is solar insolation. Keeping solar luminosity and Neptune's albedo fixed, Neptune's 0.4 bar level cannot cool below 47 K. Moreover, the Sun will brighten continuously for about 6 billion years (reaching 1.1 L$_{Sun}$ in 1.1 Gyr and 1.4 L$_{Sun}$ in 3.5 Gyr; Sackmann et al. 1993). Thus, the maximum probability of forming oceans in Neptune, while the Sun is on the main sequence, is the present probability (only 13.1$^{+5.4}_{-4.3}\%$).

As the Sun slowly (compared to Neptune's orbital period) loses about half its mass through the red giant and AGB phases, Neptune's orbit will gradually expand. Neptune will either collide with Uranus, be ejected from the Solar System, or assume a stable orbit with roughly twice its current semimajor axis (Debes and Sigurdsson 2002). This comes from angular momentum conservation:

\begin{equation}
L_{Neptune}=M_{Neptune}\sqrt{GM_{Sun}a_{Neptune}\left(1-e_{Neptune}^{2}\right)},\\
\end{equation}

\bigskip
\noindent where L is angular momentum, M is mass of each body, a is semimajor axis and e is eccentricity. As a very young white dwarf, the Sun's luminosity will be large. However, as it rapidly cools, the Sun's luminosity will decrease dramatically.

A collision would certainly mix Neptune's interior water ice into its atmosphere, and the essentially absent solar irradiance would allow the surviving planet to cool quickly. Significant cooling will also occur if Neptune is ejected from the Solar System. Thus, regardless of Neptune's eventual state, it may be free to cool down below 30 K, where its water clouds have a
 41.5 $\pm$ 4.2$\%$ probability of condensing and forming oceans. Billions of years from now, after the Sun has gone, Neptune may therefore become the only object in the Solar System with liquid water oceans.
 
\bigskip
\begin{center}
\bf{5. THE WATER-HYDROGEN-HELIUM SYSTEM}\\
\end{center}
\bigskip

We integrate the photospheric adiabat without methane, ammonia, and hydrogen sulfide to show that the conclusions above are unchanged. We set the heat capacity and latent heat due to these species to zero; therefore, this model atmosphere only contains hydrogen gas, helium, and water vapor. In Fig. 10, the bold, solid curve indicates the photospheric adiabat pinned at 59 K and containing all six species. The error bounds are given as the thin, solid curves. The 59 K photospheric adiabat containing only three species is shown as the bold, dashed curve. The error bounds are left off of the three-species adiabat for clarity. Table 2 presents the locations of all cloud bases.

The calculation with water as the only condensable still predicts a water cloud base, as opposed to an ocean surface, though the cloud base is now at lower temperature and pressure (from 663$^{+42}_{-41}$ K, 10.7$^{+8.8}_{-5.7}$ kbar to 645$^{+43}_{-28}$ K, 6.3$^{+5.2}_{-2.6}$ kbar). This effect is primarily due to the lack of a methane cloud near the photosphere, as can be seen in Fig. 10. Clouds act to steepen dP/dT, so eliminating clouds of methane and the other condensables should indeed cause the pressure at the phase transition to be lower. Since a critical ocean requires the temperature to be critical and the pressure to be supercritical, clouds of other species therefore make conditions slightly more favorable for the existence of liquid water. Indeed, a critical ocean requires less water when clouds of many species are present (X = 38.8 $\pm$ 1.4 mol-$\%$ and f = 33.0 $\pm$ 1.3 mol-$\%$, see Section 4.1) than when only water clouds are present (X = 40.4 $\pm$ 1.3 mol-$\%$ and f = 35.4$^{+1.3}_{-1.2}$ mol-$\%$).
 
\bigskip
\begin{center}
\bf{6. EXTRASOLAR HOT NEPTUNES}\\
\end{center}
\bigskip

Most extrasolar planets discovered are of order one Jupiter mass and reside less than about 1 AU from their parent stars (http://exoplanets.org); they presumably migrated many AU inward from their sites of formation (Lin et al. 1996, Boss 1996). Neptune-mass planets that have also migrated inward ( ``hot Neptunes") are beginning to be found around other stars (Santos et al. 2004, Bonfils et al. 2005, Udry et al. 2006, Lovis et al. 2006). It is reasonable to expect that many more will soon be found as technological accuracy increases. We show above that while the temperature in Neptune rapidly approaches the critical temperature with depth, the water mixing ratio reaches the deep interior value before the critical curve is reached. This is why a cloud base is reached in Neptune as opposed to an ocean surface. Thus, if Neptune at 30 AU is too hot to allow liquid water oceans to exist in its interior, then hot Neptunes at less than 1 AU must be far too hot.

We present a simplified assessment of how migration affects the possibility of liquid water oceans. We consider a family of extrasolar ice giants that have Neptune's deep interior composition and Bond albedo of 0.29, have saturated upper atmospheres, orbit stars of solar luminosity, have semimajor axes between 1 AU to 50 AU, and have photospheres that are in thermal equilibrium with their parent stars. We also assume that the atmospheres are convecting all the way down to the critical temperature and that they equilibrate to moist adiabatic states throughout the inward migration. We ignore any effects due to radiative zones in the ice giants. These assumptions may not stand up to rigorous numerical calculation; however, our goal is to present an idealized description of the effect of planetary migration on the existence of oceans. We calculate the effective planetary temperature by balancing solar heating with thermal emission (see Eq. 11). The inner semimajor axis limit, 1 AU, is chosen because the cloud base occurs at 0.4 bars: for closer semimajor axes, no cloud base would exist below a 0.4 bar photosphere. The outer choice of 50 AU is arbitrary.

As can be seen in Fig. 11, closer-in, hotter ice giants require a larger deep interior water fraction in order to have liquid water in their interiors. Thus, as Neptune-mass planets migrate inward, any ocean surface they have may evaporate into a cloud base, assuming stellar insolation is able to propagate down to this level. A Neptune-like planet with a 273 K photosphere, and therefore with liquid water droplets present in its photosphere, would need a deep interior water mixing ratio of almost 50$\%$ in order to harbor an ocean. Therefore, Neptune-like planets are very unlikely to have liquid water oceans in their interiors if water vapor is detected in their atmospheres. Conversely, the non-detection of water vapor in Neptune-like planets is actually favorable towards the existence of interior oceans. Even though photospheric water vapor should cause albedo to be different from that of Neptune, effective temperature only varies as the fourth root of albedo.

By inverting the process in Section 3.1, we calculate deep interior gas mass as a function of deep interior water mixing ratio. From Section 4.1, Neptune's minimum deep interior water mixing ratio for an ocean is f = 33.0 $\pm$ 1.3 mol-$\%$. This corresponds to a maximum deep interior gas mass of 1.40$^{+0.11}_{-0.10}$ M$_\oplus$. We assume Neptune's gas mass to be 2.0$^{+1.2}_{-0.5}$ M$_\oplus$ (see Section 3.1), which is just slightly higher than the maximum value for an ocean.

If the ``ocean planets" of Leger et al. (2004) have deep enough atmospheres, liquid water oceans can exist inside them provided the deep interior has very little gas. For example, a 6 M$_\oplus$ planet with 3 M$_\oplus$ of interior ice and a 300 K photosphere can only have an ocean if the deep interior has less than 0.117$^{+0.036}_{-0.030}$ M$_\oplus$ of gas (f $\ge$ 47.9$^{+2.9}_{-3.1}$ mol-$\%$).

\bigskip
\begin{center}
\bf{7. FUTURE WORK}\\
\end{center}
\bigskip

We have used the van der Waals equation of state in this work, and we have shown it to be accurate to only 30$\%$ in describing temperature, pressure, and composition simultaneously for T $\approx$ 450 K (see Fig. 3). The next step is to continue this work using, for example, the modified Redlich-Kwong equation of state (Redlich and Kwong 1949), which is more consistent with the phase transition curves of SF. We would also like to see whether an ocean of density 0.8 g/cm$^3$ descending from 19 kbar can successfully be incorporated into models of Neptune's density structure. To determine the conditions appropriate for cool oceans, the mixing ratio along the photospheric adiabat of water in the condensed phase must be obtained. This avenue would be useful when describing water-rich planets with less massive atmospheres. Finally, it would be beneficial to have an accurate treatment of Neptune's true photosphere with age, which will of course depend on composition. This will better address (1) whether it is possible for Neptune to cool down enough to permit liquid oceans to rain out, and (2) the length of time before this may happen.

\bigskip
\begin{center}
\bf{8. CONCLUSION}\\
\end{center}
\bigskip

Neptune's significant water content raises the interesting possibility that liquid water-hydrogen oceans, with a saturated vapor to liquid interface, exist in its interior. This liquid would be infused with over 60$\%$ hydrogen to (hydrogen + water) by mole. To be a true liquid, this ocean would have to lie at a temperature lower than the critical temperature, and a pressure higher than the critical pressure, for this composition (about 700 K and 12 kbar). There is a minimum deep interior water mixing ratio in Neptune that allows an ocean to exist. Neptune's deep interior ice mass based on density models in the literature, \begin{math} f_{interior}=26.9^{+5.2}_{-9.5} \end{math} mol-$\%$, is less than the minimum required value of f = 33.0 $\pm$ 1.3 mol-$\%$ for a critical ocean. Indeed, we find that Neptune currently has less than a 15$\%$ probability of harboring an ocean. The gravitational constraints confirm this low probability, because an extant liquid water-hydrogen ocean would be denser than measured (0.8 g/cm$^3$ instead of 0.1 to 0.3 g/cm$^3$ at the $\approx$ 15 kbar level). Thus, Neptune is both too warm and too dry for an ocean to exist at present. If the photosphere were to cool from its current 59 K to 30 K, as hydrogen gas itself begins to condense out of the atmosphere, the probability of the water clouds raining out would increase to 40$\%$.

As the Sun ages and becomes a cool white dwarf, its buffering of Neptune's atmosphere will decrease significantly. Neptune may be allowed to cool sufficiently in the ensuing billions of years for its existing water clouds to rain out. Thus, it is possible that Neptune may form liquid water oceans many billions of years from now. While terrestrial extrasolar planets with semimajor axes near 1 AU may have liquid water oceans on their surfaces, those oceans would freeze out for more distant semimajor axes. However, the inner reaches of extrasolar systems are apparently too hot for the existence of liquid water oceans in the interiors of Neptune-mass ice giants. Only the frigid conditions at many tens of AU are suitable, if the planet is watery enough, for hydrogen-rich oceans to lie at thousands of atmospheres of pressure.

\bigskip
\begin{center}
\bf{APPENDIX}\\
\end{center}
\bigskip

To derive the equation for dV/dT, we note the following thermodynamic equations:

\begin{equation}
\setcounter{equation}{1}
\renewcommand{\theequation}{A\arabic{equation}}
TdS=C_vdT+T\left(\displaystyle\frac{\partial P}{\partial T}\right)_VdV \mbox{ (Zemansky 1957, p. 245), and}
\end{equation}

\begin{equation}
\renewcommand{\theequation}{A\arabic{equation}}
TdS=C_pdT-T\left(\displaystyle\frac{\partial V}{\partial T}\right)_PdP \mbox{ (Zemansky 1957, p. 246).}
\end{equation}

\bigskip
\noindent Therefore,

\begin{equation}
\renewcommand{\theequation}{A\arabic{equation}}
C_p-C_v=-T\left(\displaystyle\frac{\partial V}{\partial T}\right)_P^2\left(\displaystyle\frac{\partial P}{\partial V}\right)_T \mbox{ (Zemansky 1957, p. 251).}
\end{equation}

\bigskip
\noindent Evaluate Eq. A3 for the van der Waals equation of state (with a and b coefficients for hydrogen gas) by taking the partial derivative of Eq. 1 with respect to V at constant P:

 \begin{equation}
\renewcommand{\theequation}{A\arabic{equation}}
\left(\displaystyle\frac{\partial V}{\partial T}\right)_P^{-1}\displaystyle\frac{R}{V-b_{H2}}-\displaystyle\frac{RT}{\left(V-b_{H2}\right)^2}+\displaystyle\frac{2a_{H2}}{V^3}=0
\end{equation}
 
 \begin{equation}
\renewcommand{\theequation}{A\arabic{equation}}
\left(\displaystyle\frac{\partial V}{\partial T}\right)_P=\displaystyle\frac{\displaystyle\frac{R}{V-b_{H2}}}{\displaystyle\frac{RT}{\left(V-b_{H2}\right)^2}-\displaystyle\frac{2a_{H2}}{V^3}}
\end{equation}
 
 \begin{equation}
\renewcommand{\theequation}{A\arabic{equation}}
\left(\displaystyle\frac{\partial P}{\partial V}\right)_T=-\displaystyle\frac{RT}{\left(V-b_{H2}\right)^2}+\displaystyle\frac{2a_{H2}}{V^3} \mbox{, so}
\end{equation}
 
 \begin{equation}
\renewcommand{\theequation}{A\arabic{equation}}
C_p-C_v=\displaystyle\frac{R^2T^2}{T\left(V-b_{H2}\right)^2}\left(\displaystyle\frac{1}{\displaystyle\frac{RT}{\left(V-b_{H2}\right)^2}-\displaystyle\frac{2a_{H2}}{V^3}}\right) \mbox{. Thus,}
\end{equation}

 \begin{equation}
\renewcommand{\theequation}{A\arabic{equation}}
C_p-C_v=\displaystyle\frac{R}{1-\displaystyle\frac{2a_{H2}\left(V-b_{H2}\right)^2}{RTV^3}},
\end{equation}
 
 \bigskip
 \noindent which is the same as Eq. 4 for one species. Adding latent heat to Eq. A1, we find

\begin{equation}
\renewcommand{\theequation}{A\arabic{equation}}
TdS=C_vdT+T\left(\displaystyle\frac{\partial P}{\partial T}\right)_VdV+\displaystyle\sum_i\left(1-f_i\right)L_idq_i=0,
\end{equation}

\bigskip
\noindent where the factor (1 - f$_i$) converts to Òper moles of mixture.Ó From the definition of q$_i$ in Eq. 3, we see that

\begin{equation}
\renewcommand{\theequation}{A\arabic{equation}}
1-f_i=\displaystyle\frac{1}{1+q_i}.
\end{equation}

\bigskip
\noindent Rearranging terms,

\begin{equation}
\renewcommand{\theequation}{A\arabic{equation}}
\left[C_v+\displaystyle\sum_i\left(\displaystyle\frac{L_i}{1+q_i}\right)\left(\displaystyle\frac{\partial q_i}{\partial T}\right)_V\right]dT+\left[T\left(\displaystyle\frac{\partial P}{\partial T}\right)_V+\displaystyle\sum_i\left(\displaystyle\frac{L_i}{1+q_i}\right)\left(\displaystyle\frac{\partial q_i}{\partial V}\right)_T\right]dV=0 \mbox{, but}
\end{equation}

\begin{equation}
\renewcommand{\theequation}{A\arabic{equation}}
T\left(\displaystyle\frac{\partial P}{\partial T}\right)_V=\displaystyle\frac{RT}{V-b_{H2}}
\end{equation}
 
 \bigskip
\noindent for a van der Waals equation of state. Thus,

\begin{equation}
\left(\displaystyle\frac{dV}{dT}\right)_S=-\left[\displaystyle\frac{C_v+\displaystyle\sum_{i}\left(\frac{L_i}{1+q_i}\right)\left(\frac{\partial q_i}{\partial T}\right)_V}{\displaystyle\frac{RT}{V-b_{H2}}+\displaystyle\sum_{i}\left(\frac{L_i}{1+q_i}\right)\left(\frac{\partial q_i}{\partial V}\right)_T}\right].\\
\end{equation}

\bigskip
\begin{center}
\bf{ACKNOWLEDGEMENTS}\\
\end{center}
\bigskip
 
We would like to thank D. J. Stevenson for a valuable debate regarding condensation in a water-hydrogen mixture. SJW would like to thank C. J. Wiktorowicz for his help in determining uncertainty along the photospheric adiabat.
 
 \bigskip
\begin{center}
\bf{REFERENCES}\\
\end{center}
\bigskip
 
\setlength{\hangindent}{0.5in}
\hangafter=1
\noindent Atreya, S. K., 1986. Atmospheres and Ionospheres of the Outer Planets and their Satellites. Springer-Verlag, New York.

\medskip
\setlength{\hangindent}{0.5in}
\hangafter=1
\noindent Atreya, S. K., Egeler, P. A., Wong, A. S. 2005. Water-ammonia ionic ocean on Uranus and Neptune-clue from tropospheric hydrogen sulfide clouds. Amer. Geophys. Union, Fall Meeting, P11A-0088  (abstract).

\medskip
\setlength{\hangindent}{0.5in}
\hangafter=1
\noindent Bonfils, X., Forveille, T., Delfosse, X., Udry, S., Mayor, M., Perrier, C., Bouchy, F., Pepe, F., Queloz,  D., Bertaux, J.-L. 2005. The HARPS search for southern extra-solar planets. VI. A Neptune-mass planet around the nearby M dwarf Gl 581. Astron. $\&$ Astrophys. 443, L15-L18.

\medskip
\setlength{\hangindent}{0.5in}
\hangafter=1
\noindent Boss, A. P. 1996. Forming a Jupiter-like companion for 51 Pegasi. Lunar. Planet. Sci. XXVII, 139-140   (abstract).

\medskip
\setlength{\hangindent}{0.5in}
\hangafter=1
\noindent Burgdorf, M., Orton, G. S., Davis, G. R., Sidher, S. D., Feuchtgruber, H., Griffin, M. J., Swinyard, B. M.   2003. NeptuneÕs far-infrared spectrum from the ISO long-wavelength and short-wavelength spectrometers. Icarus 164, 244-253.

\medskip
\setlength{\hangindent}{0.5in}
\hangafter=1
\noindent Debes, J. H., Sigurdsson, S. 2002. Are there unstable planetary systems around white dwarfs? Astrophys.   J. 572, 556-565.

\medskip
\setlength{\hangindent}{0.5in}
\hangafter=1
\noindent Emanuel, K. A. 1994. Atmospheric Convection. Oxford University Press, New York.

\medskip
\setlength{\hangindent}{0.5in}
\hangafter=1
\noindent Fishbane, P. M., Gasiorowicz, S. G., Thornton, S. T., 2005. Physics for Scientists and Engineers,   Prentice-Hall, Upper Saddle River.

\medskip
\setlength{\hangindent}{0.5in}
\hangafter=1
\noindent Gautier, D., Conrath, B. J., Owen, T., de Pater, I., Atreya, S. K. 1995. The troposphere of Neptune.   In: Cruikshank, D. P. (Ed.), Neptune and Triton. University of Arizona Press, Tucson, pp. 547-611.

\medskip
\setlength{\hangindent}{0.5in}
\hangafter=1
\noindent Giauque, W. F., Blue, R. W. 1936. Hydrogen sulfide. The heat capacity and vapor pressure of solid   and liquid. The heat of vaporization. A comparison of thermodynamic and spectroscopic values of the entropy. J. Am. Chem. Soc. 58, 831-837.

\medskip
\setlength{\hangindent}{0.5in}
\hangafter=1
\noindent Hubbard, W. B. 1972. Statistical mechanics of light elements at high pressure. II. Hydrogen and helium   alloys. Astrophys. J. 176, 525-531.

\medskip
\setlength{\hangindent}{0.5in}
\hangafter=1
\noindent Hubbard, W. B., Podolak, M., Stevenson, D. J. 1995. The interior of Neptune. In: Cruikshank, D. P.   (Ed.), Neptune and Triton. University of Arizona Press, Tucson, pp. 109-138.

\medskip
\setlength{\hangindent}{0.5in}
\hangafter=1
\noindent Hubbard, W. B. 1999. Interiors of the giant planets. In: Beatty, J. K., Petersen, C. C., Chaikin, A.   (Eds.), The New Solar System. Sky Publishing Co., Cambridge, MA, pp. 193-200.

\medskip
\setlength{\hangindent}{0.5in}
\hangafter=1
\noindent International Critical Tables. 1928. McGraw-Hill Book, New York.

\medskip
\setlength{\hangindent}{0.5in}
\hangafter=1
\noindent Karwat, E. 1924. Der Dampfdruck des festen Chlorwasserstoffs, Methans und Ammoniaks. Z. Phys.   Chem. St\ ''{o}chiomet. Verwandtschaftsr. 112, 486-490.

\medskip
\setlength{\hangindent}{0.5in}
\hangafter=1
\noindent L\'{e}ger, A., 11 colleagues 2004. A new family of planets? ÒOcean-Planets.Ó Icarus 169, 499-504.

\medskip
\setlength{\hangindent}{0.5in}
\hangafter=1
\noindent Lin, D. N. C., Bodenheimer, P., Richardson, D. C. 1996. Orbital migration of the planetary companion   of 51 Pegasi to its present location. Nature 380, 606-607.

\medskip
\setlength{\hangindent}{0.5in}
\hangafter=1
\noindent Lindal, G. F. 1992. The atmosphere of Neptune: an analysis of radio occultation data acquired with   Voyager 2. Astron. J. 103, 967-982.

\medskip
\setlength{\hangindent}{0.5in}
\hangafter=1
\noindent Lodders, K., Fegley, B., Jr. 1994. The origin of carbon monoxide in NeptuneÕs atmosphere. Icarus 112,   368-375.

\medskip
\setlength{\hangindent}{0.5in}
\hangafter=1
\noindent Lovis, C., 14 colleagues 2006. An extrasolar planetary system with three Neptune-mass planets. Nature   441, 305-309.

\medskip
\setlength{\hangindent}{0.5in}
\hangafter=1
\noindent Marley, M. S., G\'{o}mez, P., Podolak, M. 1995. Monte Carlo interior models for Uranus and Neptune. J.   Geophys. Res. 100, 23349-23353.

\medskip
\setlength{\hangindent}{0.5in}
\hangafter=1
\noindent National Institute of Standards and Technology: http://webbook.nist.gov/chemistry/name-ser.html

\medskip
\setlength{\hangindent}{0.5in}
\hangafter=1
\noindent Osborne, N. S. and van Dusen, M. S. 1918. Latent heat of vaporization of ammonia. J. Am. Chem.   Soc., 40, A14-A25.

\medskip
\setlength{\hangindent}{0.5in}
\hangafter=1
\noindent Podolak, M., Reynolds, R. T. 1984. Consistency tests of cosmogonic theories from models of Uranus and   Neptune. Icarus 57, 102-111.

\medskip
\setlength{\hangindent}{0.5in}
\hangafter=1
\noindent Podolak, M., Hubbard, W. B., Stevenson, D. J. 1991. Models of UranusÕ interior and magnetic field. In:   Bergstralh, J., Minor, E., Matthews, M. S. (Eds.), Uranus. University of Arizona Press, Tucson, pp. 29-61.

\medskip
\setlength{\hangindent}{0.5in}
\hangafter=1
\noindent Podolak, M., Podolak, J. I., Marley, M. S. 2000. Further investigations of random models of Uranus and   Neptune. Planet. Space Sci. 48, 143-151.

\medskip
\setlength{\hangindent}{0.5in}
\hangafter=1
\noindent Redlich, O., Kwong, J. N. S. 1949. On the thermodynamics of solutions. V. An equation of state. Chem.   Rev. 44, 233-244.

\medskip
\setlength{\hangindent}{0.5in}
\hangafter=1
\noindent Sackmann, I.-J., Boothroyd, A. I., Kraemer, K. E. 1993. Our Sun. III. Present and Future. Astrophys.   J. 418, 457-468.

\medskip
\setlength{\hangindent}{0.5in}
\hangafter=1
\noindent Salby, M. L. 1996. Atmospheric Physics. Academic Press, San Diego.

\medskip
\setlength{\hangindent}{0.5in}
\hangafter=1
\noindent Santos, N. C., 15 colleagues 2004. The HARPS search for southern extra-solar planets. II. A 14 Earth-  masses Exoplanet around $\mu$ Arae. Astron. $\&$ Astrophys. 426, L19-L23.

\medskip
\setlength{\hangindent}{0.5in}
\hangafter=1
\noindent Seward, T. M., Franck, E. U. 1981. The system hydrogen Ð water up to 440$^\circ$C and 2500 bar pressure.   Ber. Bunsen Phys. Chem. 85, 2-7. 

\medskip
\setlength{\hangindent}{0.5in}
\hangafter=1
\noindent Stanley, S., Bloxham, J. 2004. Convective-region geometry as the cause of UranusÕ and NeptuneÕs unusual   magnetic fields. Nature 428, 151-153.

\medskip
\setlength{\hangindent}{0.5in}
\hangafter=1
\noindent Udry, S., Mayor, M., Benz, W., Bertaux, J.-L., Bouchy, F., Lovis, C., Mordasini, C., Pepe, F., Queloz,   D., Sivan, J.-P. 2006. The HARPS search for southern extra-solar planets. V. A 14 Earth-masses planet orbiting HD 4308. Astron. $\&$ Astrophys. 447, 361-367.

\medskip
\setlength{\hangindent}{0.5in}
\hangafter=1
\noindent Vorholz, J., Rumpf, B., Maurer, G. 2002. Prediction of the vapor-liquid phase equilibrium of hydrogen   sulfide and the binary system water-hydrogen sulfide by molecular simulation. Phys. Chem. Chem. Phys. 4, 4449-4457.

\medskip
\setlength{\hangindent}{0.5in}
\hangafter=1
\noindent Wagner, W., Pruss, A. 1993. International equations for the saturation properties of ordinary water   substance --- revised according to the international temperature scale of 1990. J. Phys. Chem. Ref. Data 22, 783-787.

\medskip
\setlength{\hangindent}{0.5in}
\hangafter=1
\noindent Zemansky, M. W. 1957. Heat and thermodynamics, McGraw-Hill Book Company, Inc., New York.

\medskip
\setlength{\hangindent}{0.5in}
\hangafter=1
\noindent Ziegler, W. T. 1959. The vapor pressures of some hydrocarbons in the liquid and solid state at low   temperatures. Natl. Bur. Stand. Tech. Notes 6038, Washington, DC.

\pagebreak
\begin{center}
{\bf{Table 1}\\
\bf{Deep Interior Mixing Ratios}}

\bigskip
\begin{tabular}{cc}
\qquad \qquad \enspace {Ice and gas separate}&{Ice and gas combined}
\end{tabular}

\smallskip
\begin{tabular}{|c|c|c|c|c|}
\hline
Species&{By mole}&{By mass}&{By mole}&{By mass}\\
\hline
{H$_2$}&{81.0$\%$}&{68.1$\%$}&{45.8$\%$}&{10.2$\%$}\\
{He}&{19.0$\%$}&{31.9$\%$}&{10.8$\%$}&{4.8$\%$}\\
{H$_2$O}&{61.7$\%$}&{63.3$\%$}&{\bf{26.9$\%$}}&{53.8$\%$}\\
{CH$_4$}&{28.8$\%$}&{26.3$\%$}&{12.5$\%$}&{22.3$\%$}\\
{NH$_3$}&{8.1$\%$}&{7.9$\%$}&{3.5$\%$}&{6.7$\%$}\\
{H$_2$S}&{1.3$\%$}&{2.6$\%$}&{0.6$\%$}&{2.2$\%$}\\
\hline
\end{tabular}

\bigskip
\bigskip

\bigskip
{\bf{Table 2}\\
\bf{Cloud Bases}}

\bigskip
 \doublespacing
\begin{tabular}{|c|c|c|c|c|}
\hline
Species&{One condensable}&{Four condensables}\\
\hline
{CH$_4$}&{---}&{$83.6^{+2.2}_{-3.1}$ K, $1.78^{+0.31}_{-0.34}$ bar}\\
{H$_2$S}&{---}&{$157.3^{+3.2}_{-4.4}$ K, $14.2^{+2.2}_{-2.4}$ bar}\\
{NH$_3$}&{---}&{$215.8^{+6.2}_{-8.4}$ K, $43.0^{+8.9}_{-9.5}$ bar}\\
{H$_2$O}&{$645^{+43}_{-28}$ K, $6.3^{+5.2}_{-2.6}$ kbar}&{$663^{+42}_{-41}$ K, $10.7^{+8.8}_{-5.7}$ kbar}\\
\hline
\end{tabular}

\pagebreak
\bf{FIGURE CAPTIONS}\\
\end{center}
\bigskip

\noindent Figure 1. Temperature-entropy curve for pure water. Note that an adiabat is a vertical line in this diagram and that temperature is shown increasing downward.

\bigskip
\noindent Figure 2. Photospheric adiabat overlying SF phase boundaries. Phase boundaries are thin lines, and numbers on each boundary denote hydrogen to (water + hydrogen) mixing ratio. The photospheric adiabat is given as the thick line and follows, for the most part, the 90 mol-$\%$ phase boundary. However, it can be seen that the water mixing ratio in the gas phase {\it{increases}} along the photospheric adiabat as temperature is increased (hydrogen mixing ratio {\it{decreases}}).

\bigskip
\noindent Figure 3. Photospheric adiabat uncertainty. We sample nine (T, P, f) points along two experimentally determined phase boundaries (f = 10 mol-$\%$ and 40 mol-$\%$) of SF. Using T and f for each point, we calculate the volume from the photospheric adiabat using Eq. 5. We then calculate the volume from experiment using Eq. 1. The ratio of these volumes gives a correction factor which is then applied to the photospheric adiabatÕs slope. We assume that ideal gas conditions at 273 K imply the uncertainty here to be zero.

\bigskip
\noindent Figure 4. Pressure-temperature profile along the photospheric adiabat. The van der Waals, photospheric adiabat containing H$_2$, He, water vapor, CH$_4$, NH$_3$, and H$_2$S is pinned at 59 K, 0.4 bar and extends to the critical temperature for pure water (647 K). The photospheric adiabat reaches this temperature at $8.8^{+1.8}_{-3.1}$ kbar.

\bigskip
\noindent Figure 5. Critical curve: temperature vs. X. Data from SF are fit by a piecewise cubic interpolating Hermite polynomial.

\bigskip
\noindent Figure 6. Critical curve: pressure vs. X. Data from SF are fit by a piecewise cubic interpolating Hermite polynomial.

\bigskip
\noindent Figure 7. Minimum water mixing ratio for an ocean. The photospheric adiabat, extrapolated for T $>$ 647 K, and the critical curve are shown. Dashed curves indicate 1$\sigma$ errors on each profile. The two curves intersect at X = 38.8 $\pm$ 1.4 mol-$\%$. This represents the minimum water to (water + hydrogen) mixing ratio necessary to support an ocean under the current, 59 K Neptunian photosphere. Since the deep interior f corresponds to X = 32$^{+6}_{-11}$ mol-$\%$ at an ocean, Neptune is too dry to harbor oceans.

\bigskip
\noindent Figure 8. Density along the photospheric adiabat overlying density models. The thick, linear feature in the lower left (from 10$^{-4}$ to 10$^{-2}$ Mbar) represents pressure versus density along the photospheric adiabat. 1$\sigma$ error bounds are roughly the thickness of the line. The cross at about 10$^{-2}$ Mbar and about 1 g/cm$^3$ shows the location of the critical ocean, under a 59 K photosphere, in pressure-density space. This figure is modified from Fig. 5 of Hubbard et al. (1995), and thin curves (solid, dashed, and dot-dashed) are different density models in that paper. Numbered, dotted curves represent adiabats from that paper labeled with varying ice mass fractions. These adiabats lie in ice giants composed only of hydrogen, helium, and ice. Note that the photospheric adiabat lies within the range of density models while the ocean surface is far outside them. Thus, we conclude not only that Neptune has no extant liquid water-hydrogen ocean, but also that the photospheric adiabat is consistent with Voyager data.

\bigskip
\noindent Figure 9. Ocean probability. With a 30 K photosphere, Neptune would have a 41.5 $\pm$ 4.2$\%$ probability of having an ocean, while its current 59 K photosphere only has a probability of 13.1$^{+5.4}_{-4.3}\%$ of overlying an ocean. This is because liquid water preferentially exists at low entropy and thus low atmospheric temperature.

\bigskip
\noindent Figure 10. Pressure-temperature profiles along photospheric adiabats with different condensable species. The solid curves represent the photospheric adiabat (and error bounds) with all species condensing (water, CH$_4$, H$_2$S, and NH$_3$), whereas the dashed curve is a photospheric adiabat with only water condensing. For clarity, error bounds are only shown for the adiabat with all species condensing. All photospheric adiabats are pinned at 59 K. Note that more condensation causes a photospheric adiabat to run colder at depth, thus increasing the probability of an oceanÕs existence.

\bigskip
\noindent Figure 11. Minimum deep interior water mixing ratio necessary for ocean existence. As Neptune-like planets migrate inward, and planetary effective temperature increases, oceans that once existed may boil away. The solid curve represents the minimum planetary f necessary for an ocean to exist in an ice giant at a given semimajor axis a. The effective temperature T$_e$ assumes a Neptune-like Bond albedo of 0.29. The large, required f at effective temperatures higher than 200 K casts serious doubt on the idea that hot Neptunes (a $<$ 1 AU) can harbor liquid water oceans. Neptune itself lies at the filled circle in the lower right of the diagram, and it is too dry (finterior Å 27 mol-$\%$) to permit the existence of oceans.

 \end{document}